\documentclass[12pt,twoside]{article} 
\usepackage[english]{babel}
\usepackage{graphicx}
\usepackage{amsmath,amssymb,amsfonts}
\usepackage{mathtools,slashed,braket,tensor,accents}
\usepackage[makeroom]{cancel}
\usepackage{ifthen}
\usepackage{setspace}
\usepackage[utf8]{inputenc}
\usepackage{soul}
\usepackage[misc]{ifsym}
\usepackage[normalem]{ulem}
\usepackage{wrapfig}
\usepackage{lipsum} % For generating dummy text, can be removed in your actual document
\usepackage{float}
\usepackage{subfloat}
\usepackage[export]{adjustbox}
\usepackage{tikz}
\usepackage[textsize=tiny,color=red]{todonotes}

 \usepackage{kotex}%Korean
\usepackage{CJKutf8}
\usepackage[font=small, labelfont=bf]{caption}
\usepackage{subcaption, comment}
 \usepackage[]{titlesec} %you can choose e.g. [indentafter]
 \titleformat{name=\section}[block]{}{\thesection.}{0.8em}{\centering\bfseries}
 \titleformat{name=\subsection}[runin]{}{\thesubsection.}{0.5em}{\bfseries}[.]
\titleformat{name=\subsubsection}[runin]{}{\thetitle.}{0.5em}{\itshape}[.]

\usepackage[top=2.5cm, bottom=2.5cm, left=2.5cm, right=2.5cm]{geometry} 
\usepackage{setspace} 
\usepackage{color}
\definecolor{lighterblue}{RGB}{120,140,250}
\definecolor{lightblue}{RGB}{50,120,150}
\definecolor{darkred}{RGB}{190,0,0}
\definecolor{darkerred}{RGB}{170,0,0}
\definecolor{darkblue}{rgb}{0.0, 0.0, 0.5}
\definecolor{amber}{rgb}{1.0, 0.75,0.0}
\definecolor{arsenic}{RGB}{50,50,90}
\definecolor{goodgreen}{rgb}{0.0, 0.55, 0.2}
\definecolor{darkorange}{RGB}{255,100,0}
\definecolor{antiquefuchsia}{rgb}{1.5, 0.1, 0.5}
\definecolor{groen}{RGB}{50,156,105}
\definecolor{paarsblauw}{RGB}{92,86,209} %#5C56D1
 
\usepackage[colorlinks=true,
            linkcolor=antiquefuchsia,  % groen,
            urlcolor=darkblue,
            citecolor=paarsblauw
            ]{hyperref}      

 \usepackage[
 style=ext-authoryear-comp,
 maxcitenames=3,
 sorting=ynt,
 uniquename=false,
 backend=biber,
 uniquelist=false,
 articlein=true,
 sortcites=true,
 natbib=true,
 doi=false,
 url=false,
 isbn=false]{biblatex}
\DeclareNameAlias{sortname}{last-first}
\makeatletter
\renewcommand\hyper@natlinkbreak[2]{#1}
\makeatother
 \renewbibmacro{in:}{}
\DeclareNameAlias{sortname}{last-first}

 \DefineBibliographyStrings{english}{%
  byeditor = {Edited by},
}

\DefineBibliographyStrings{english}{%
  backrefpage = {cited on},
  backrefpages = {cited on},
}

 \newbibmacro{string+doiurlisbn}[1]{%
  \iffieldundef{doi}{%
    \iffieldundef{url}{%
      \iffieldundef{isbn}{%
        \iffieldundef{issn}{%
          #1%
        }{%
          \href{http://books.google.com/books?vid=ISSN\thefield{issn}}{#1}%
        }%
      }{%
        \href{http://books.google.com/books?vid=ISBN\thefield{isbn}}{#1}%
      }%
    }{%
      \href{\thefield{url}}{#1}%
    }%
  }{%
    \href{http://dx.doi.org/\thefield{doi}}{#1}%
  }%
}

\DeclareFieldFormat{title}{\usebibmacro{string+doiurlisbn}{\mkbibemph{#1}}}
\DeclareFieldFormat[article,incollection]{title}%
    {\usebibmacro{string+doiurlisbn}{\mkbibquote{#1}}}
  
\usepackage{tensind}
  \tensordelimiter{?}

\usepackage{thmtools} % Use 'restatable' to restate theorem in Appendix
\usepackage{thm-restate}

\newsavebox\MBox

\numberwithin{equation}{section}

\usepackage{xparse}
\usepackage{cleveref}

\usepackage{placeins}
\usepackage{float}

\usepackage[nodayofweek,level]{datetime}
\newcommand{\mydate}{\formatdate{17}{09}{2026}}
\date{\mydate}

\begin{document}
% \title{\textsc{Conventionalism in general relativity: \\ 
% % Reichenbach's 
% Weatherall $\&$ Manchak's proof against theorem θ in context }}
\title{The spacetime/matter-dichotomy in the early universe: 
conventionalist temptation through idealization}
 \author{Ruward Mulder\footnote{\it University of California, Irvine \it Department of Logic and Philosophy of Science, 739 Social Science Tower, 
 Irvine California, CA 92697-5100. \href{mailto:ramulder@uci.edu}{ramulder@uci.edu}.}%\footnote{\it University of Geneva, \it Department of Philosophy, 5 rue de Candolle, 1211, Geneva 4, Switzerland.}
 }
 \maketitle
\vspace{-0.5cm}
\setstretch{0.9}
% \vspace{-1cm}
 \begin{abstract}
\noindent {\footnotesize 
Can the boundary between ``spacetime" and ``matter" be redrawn merely by changing mathematical representation? 
A striking case concerns the formal equivalences between $f(R)$-gravity and its Brans-Dicke scalar-tensor formulations---the (in)famous Jordan and Einstein frames. 
Under standard GR textbook coordinations, the theory's additional scalar degree of freedom is interpreted as representing spacetime curvature; 
after transformation to the Einstein frame, however, that same degree of freedom appears as an ordinary scalar matter field. 
This apparent equivalence has invoked conventionalist temptations to reject the conceptual spacetime/matter-dichotomy, especially in inflationary cosmology,  where that scalar dominates observables.  
After classifying the conventionalist positions and reconstructing the formal relations between $f(R)$-gravity and the Jordan and Einstein frames, this chapter first rejects the claim (Dürr 2021) that the scalar field should be eliminated on grounds of parsimony. 
It is then pointed out that the deeper problem is   not ontological extravagance but rather incompleteness through idealization. 
Exact equivalence holds only in an isolated ``pure-gravity" sector,
 but once known matter fields are restored -- illustrated here via Higgs sector completion -- using standard heuristics of canonical normalization and minimal coupling, 
the putative equivalence crumbles, giving way to approximation (in Norton's (2012) sense). 
%Thus, idealization can create a misleading appearance of conventionality, and that formal equivalence has ontological or confirmatory force only when it remains stable under relevant de-idealization.  
 The general lesson is that
model-building practices often employ idealizations that obscure ontologically salient features, making the viability of conventionalism about conceptual dichotomies merely scope relative. %, should be evaluated only for de-idealized, in the sense of \textit{completed}, theories.
}
  \end{abstract}
  
% \setstretch{0.5}
% % \vspace{-0.5em}
% \noindent{\scriptsize\textbf{Assumed background:} basic general relativity and variational principles; 
%  basic philosophy of science.} 

% \vspace{0.2cm}
% \noindent {\scriptsize \textbf{Keywords:} spacetime/matter-dichotomy, conventionalism, $f(R)$-gravity, Brans-Dicke, Jordan frame, Einstein frame, Legendre transformation, conformal transformation, idealization, approximation, conceptual engineering,  minimal coupling, Brown's moral, Starobinsky inflation.}
% % \vspace{0.75em}

 \renewcommand{\baselinestretch}{0.6} 
% \clearpage
% Table of Contents%
  \thispagestyle{empty}
 \renewcommand{\contentsname}{Table of contents}
  % \addcontentsline{toc}{section}{\protect\numberline{}Table of contents}
 % \footnotesize{
  %\tiny{
  \scriptsize{
{\hypersetup{linkcolor=black} \textsf{\tableofcontents}}
}

 \thispagestyle{empty}
\clearpage 
 \setcounter{page}{1}

% \clearpage
\normalsize{ }
% \renewcommand{\baselinestretch}{1.25} 
% \setstretch{1.25}
\setstretch{1.0}
% \pagenumbering{arabic}

% \section{}

\section{Introduction} \label{sec:intro}

The spacetime/matter-dichotomy is the conceptual distinction  between that what is represented by ``spacetime" structure of a theory from that which is represented by its ``matter" content, 
 dividing the contained from that which contains.     
Conventionalism about the spacetime/matter-dichotomy, then, is the view that this dividing  line is not uniquely fixed by nature.  
 Rather, there are multiple empirically equivalent ways to choose how that line be drawn, 
 with that choice itself not amenable to empirical justification---that is, a convention.

If viable, such conventionalism would withhold theory-independent truth values for claims that physical structures are ``spacetime-like" or ``matter-like", 
for these are not merely claims about observable phenomena (which may remain invariant across alternative frameworks) 
but claims about how the theory's structures should be classified within its physical ontology. 
Such claims may be analytically true or false if construed relative to a conventionally chosen framework (`true or false by convention'). 
 But if construed absolutely -- as claims about a framework-independent dichotomy instantiated by the world -- they lack truth values: % As far as the theory under consideration goes, 
  there would be no facts about spacetime, and no fact of the matter about matter.

Is such conventionalism viable for theories today,   
or is the dichotomy stable? 
To answer, recall the requirements for conventionalism more generally.  
Roughly, conventionalism requires multiple models, either within the same theory or between theories (\textit{given} conceptual comparability), that can represent the same phenomena but do so in conceptually conflicting ways.  
  Less roughly (cf.~\cite[Ch.~VII]{Friedman1983Foundations}; %\cite{Sklar2004Spacetime&Conventionalism} 
  \cite[pp.~88–101, 117–119, 128–129]{Sklar1974SpaceTimeSpacetime};  \cite[pp.~3--10, 39--79]{Ben-Menahem2006-BENC-2}; \cite[pp.~24--25]{Duerr2021-DUETIA}), conventionalism requires:  
\begin{itemize}
    \item[(i)] \textbf{Coordination}. A conceptual framework that provides coordinations for each relevant model with physical phenomena, i.e., that fixes a representational mapping from mathematical structures in the model to phenomena in the world; 
    \item[(ii)] \textbf{Formal equivalence}. Some mathematical structure of alternative models is coordinated, via (i), in an empirically indistinguishable way, and this is backed up formally by the availability of mutual translation (or otherwise formal equivalence in relevant respects) between those structures;  
    \item[(iii)] \textbf{Conceptual conflict}. A framework-induced conflict between those  alternative models: 
    under their respective coordinations specified in (i), 
    their corresponding formal structures are assigned mutually exclusive representational roles.  
    Such conflict can be~(iiia) context-relative, when the same structure cannot represent both concepts at once, yet could switch roles across contexts, 
        or~(iiib) absolute, when representing one concept excludes the other wholesale
    , so that a structure represents (in the words of the editors) 
        either one or the other, never both, never neither.
    \item[(iv)] \textbf{Non-trivial semantics}. Arguably, one may demand the further negative condition that the alternatives must be non-trivial 
    in that they must amount to more than mere semantic reconstruals 
    or notational repartitionings of  each other's content.
    \end{itemize}
Thus, alternatives must be more than merely different formal expressions (ii) or trivial semantic reconstruals (iv) because of a mutual contradiction arising from their individual coordinations (i)---call this \textit{genuine underdetermination} as opposed to mere \textit{formal underdetermination} \citep{MulderForthcomingGeometricTrinity}. 
      Conventionalists embrace equivalence (ii) and seek to resolve the conflict (iii) by relaxing coordination (i) in a non-trivial way (iv),
    by deleting those representational maps in (i) that are responsible for contradictions, leaving  formal underdetermination only. 
     In this sense, conventionalists are selective realists \citep[cf.]{DurrRead2024}.

   Textbook geometrical conventionalism \citep[e.g.][pp.~80--99]{Ben-Menahem2006-BENC-2} illustrates requirements (i)--(iv). This view  denies that geometric differences between alternative models correspond to factual differences. Its coordination (i) identifies physical geometry, represented by the metric, through the behaviour of rigid bodies, while coordinating universal effects as non-geometrical yet physically salient. 
 The formal basis (ii) lies in Poincaré's equivalence proofs of spaces of constant curvature, used in Reichenbach's particular construal (theorem $\theta$) to represent differences in physical geometry as the literal difference between the metrics of rival models. 
 (Reichenbach's claim is stronger than conventionalism requires: Reichenbach maintains that \textit{any} metric can perform the same empirically indistinguishable role, whereas the existence of \textit{one} distinct alternative would suffice \citep{Mulder2025-MULCIG}.) 
 The conceptual conflict (iii) arises because structure represented as geometry in one model is represented as non-geometrical in another. 
 Finally, Reichenbach's version arguably fails (iv), as his operationalism threatens to collapse empirical equivalence into theoretical equivalence: his semantics offers few resources to make these alternatives non-trivially distinct, yielding \textit{trivial semantic holism} (\cite{Roberts2026-ROBTCO-74,DandbBM2022}; cf.~\cite[pp. 99–101]{Sklar1974SpaceTimeSpacetime}).

Crucially, the viability of conventionalism is \textit{theory-dependent}. One can simultaneously regard the spacetime/matter-dichotomy as factual in the Newtonian and conventional in the relativistic context.  
Indeed, arguments for or against are formulated only after the relevant theory  is specified---aplenty in the current volume: classical electrodynamics, Newtonian gravity, QFT on curved backgrounds, spin-2 gravity, supergravity, etc.

Thus I propose a classifying tool: reframe  controversy over the spacetime/matter-dichotomy along two axes: first, specify the theory (or class of theories) under consideration, and, second, assess
whether  (i)-(iv) should be accepted, challenged, or revised. 
  
 Take  general relativity (GR). 
 The usual coordination (i) for ``matter" and ``spacetime" 
 -- call it the \textit{GR textbook coordination}  since it is ingrained in us through textbook terminology -- 
  seems simple enough at first sight: 
\textit{spacetime} is represented by the metric tensor on a differentiable manifold, along with its implicitly defined Levi-Civita connection; matter fields are those that enter into the stress-energy tensor, such as the electromagnetic or Higgs gauge potentials, and that stress-energy tensor is conserved using the covariant derivative of that same Levi-Civita connection. 
Furthermore, GR textbooks typically begin with ``vacuum" models, in which that stress-energy tensor is set to zero, so that the right-hand side of Einstein's equation vanishes, and treat solutions of this equation as representing ``empty" (i.e., \textit{devoid of matter}) possible worlds, e.g., Minkowski, Schwarzschild, pp waves, gravitational waves, and (depending on one's classification of the cosmological constant, see below) de Sitter spacetimes. 
Thus, by first establishing what spacetime is, one obtains (assuming mutual exclusivity) a complementary concept of matter as that which spacetime is not.

Dissenting voices abound: behold this volume. 
\citet{ShiChuaForthcomingMatterWithMatter} press on (i): stress-energy and non-local mass-energy notions, they argue, are not classified independently of spacetime structure, so ``matter'' in GR is  context-sensitive. %; hence the textbook inference from `vacuum'' to ``devoid of matter'' is unstable unless one simply stipulates that such degrees of freedom are not matter. 
  The cosmological constant $\Lambda$ can be treated  spacetime \citep{Baker2005-BAKSSA}, but \citet{ZitoMcCoyForthcomingAcceleratedExpansion} classify it as material, while \citet{MartensForthcomingLateTimeExpansion} remains uncommitted, taking seriously also a third ``constant of nature" category. 
   \citet{ShiForthcomingNothingReallyMatters} and \citet{DoboszewskiLuoForthcomingBlackHoles} raise analogous concerns about whether models of gravitational waves or black holes should be seen as  representing  only spacetime, given that   physics practice often regards them as causally efficacious and massive.  
\citet{LinnemannForthcomingSpin2Gravity} addresses a conflict (iii) for spin-2 gravity,  equivalent to the sector of GR in which light-cone compatibility holds (every $g$-causal direction lies inside the $\eta$-cone, implying global hyperbolicity), thereby satisfying (ii), so that the same formalism admits both   geometric and   field-on-a-background descriptions.

All the above press on (i) and/or (iii). 
 The rest of this chapter, instead,  
 considers (ii), 
bracketing the above objections and taking the \textit{GR textbook coordination} as given. 
The theory under consideration is $f(R)$\textit{-gravity}.    
% My concern is instead with condition (ii), and the case I consider is metric \textit{$f(R)$-gravity}. 
 This is a class of theories that modify GR by replacing the Einstein-Hilbert action -- linear in the Ricci scalar $R$ -- with a more general function $f(R)$. 
These are not merely formal curiosities, 
but routinely studied in cosmological models of  early-universe inflation and late-time acceleration  \citep[cf.][]{Inflationaris,SmeenkWeatherall2024DarkEnergyModifiedGravity}.   % - discuss why f(R) theory was introduced. Jacobson: For Starobinsky that was not the motivation. Rather, it's that in QFT such terms are generated by “radiative corrections” and vacuum effect, and have to be tuned to zero to be eliminated.
  Best known is Starobinsky inflation, which contains, next to the linear term, an $R^2$-term %quadratic in $R$ 
\citep{Starobinsky1980NewType};  
it remains a top-tier candidate according to recent data DESI and PLANCK data (\cite{Aghanim2020PlanckOverview,DESI2025}; cf.~\cite{FerreiraElisaBAOCMB,Wolf2025NonMinimal}).

$f(R)$-theories put pressure on the spacetime/matter-dichotomy.
In the original formulation it encoded an extra massive scalar degree of freedom in nonlinear curvature terms, which thus appears spacetime-like; after a Legendre transformation (\S\ref{sec:equivJFEF}) to the \textit{Jordan frame} (JF), followed by a conformal transformation and field redefinition, it appears in the \textit{Einstein frame} (EF) as a canonical scalar field with a potential, just like ordinary matter. 
The conventionalist temptation is therefore immediate: 
under the GR textbook coordination (i), 
formally equivalent formulations (ii) 
assign the same degree of freedom to spacetime in one representation and to matter in another, 
leading to a conceptual conflict (iii). 
%Hence, apparently, the spacetime/matter-dichotomy is representation-dependent.
 % In the Starobinsky case, this is the familiar scalaron with a plateau potential, providing a suitable mechanism for inflation.
   
 In \S\ref{sec:reviewconvtemptation}, I review some physicists and philosophers engaging this conventionalist temptation prompted by $f(R)$-gravity. 
% mention \cite{FerreiroFleurenMartens} first, who consider conventionalism and/or functionalism? 
 % Where \citet{FerreiroFleurenMartens} show in detail how the dichotomy is challenged, and indicate possible philosophical attitudes towards it, 
 In \S\ref{sec:againstparsimony}, I resist a main argument of \citep{Duerr2021-DUETIA} 
 that the spacetime interpretation is \textit{privileged} 
 because the Brans-Dicke scalar is dynamically reducible to the metric tensor and therefore, on grounds of parsimony, ontologically eliminable. 
I argue the scalar is not unparsimonious, but a formal explication of the massive scalar degree of freedom already present in the original metric dynamics. %Parsimony over variables should not be confused with parsimony over physical degrees of freedom.
 Nevertheless, I agree that the conventionalist temptation ought to be resisted by undermining (ii),  though not for reasons of parsimony. 
The reason (\S\ref{sec:conventionalistbreaksdown}) I give is that (ii)  holds only in an idealized ``purely gravitational" sector, where idealization is understood in John Norton's (\cite{Norton2012Idealization}) sense, explained in \S\ref{sec:idealapprox}.  
Stripped of known matter sectors -- numerically justified by modeling practices, certainly in the case of inflation -- it is easy to mistake the model for the target. % treated as though it represented the complete physical system. 
When completing the theory by adding matter we know to exist, here illustrated via Higgs sector completion, the equality collapses into mere approximation. 
 % coupling it minimally and canonically on both sides of the $f(R)$--EF equivalence  breaks the equality straightforwardly. 
 The general lesson is not that spacetime and matter are conventional, but that model-building idealizations can make conventionalism about a conceptual dichotomy look better motivated than it is.

 \section{Formalism: Starobinsky inflation, $f(R)$-gravity, Brans-Dicke theories, and dynamical equivalence of Einstein and Jordan frames} \label{sec:equivJFEF}

 Start with the ``vacuum" or ``pure gravity" sector of $f(R)$-gravity, and further assume ``metric $f(R)$-gravity" (i.e. the affine-connection is torsion-free and metric-compatible), which modifies Einstein-Hilbert gravity (I follow \cite{SotiriouFaraoni2010FR,
AmendolaTsujikawa2010DarkEnergy,
FaraoniGunzigNardone1999ConformalTransformations,
Quiros2019ScalarTensorTheories,Mukhanov2005PhysicalFoundations,CliftonFerreiraPadillaSkordis2012}):
\begin{equation}
S_{f(R)}[g]
=
\frac{M_P^2}{2} \int d^4x\sqrt{-g} f(R),
\label{Eqf(R)}
\end{equation}
with $M_P$ (later identified with) the reduced Planck mass and $f(R)$ some function of the Ricci scalar (of the Levi-Civita connection defined by $g_{ab}$). The \textit{modified} Einstein equations become
\begin{equation}\label{EqfREOM}
f'(R) R_{ab}
-\frac{1}{2}f(R) g_{ab}
+\left(g_{ab}\Box-\nabla_a\nabla_b\right)f'(R)
=0, 
\qquad
f'(R)\equiv df/dR, 
\end{equation}
for $\Box \equiv g^{ab}\nabla_a\nabla_b$ the d'Alembertian (associated with $g_{ab}$).  
Compared to the two massless degrees of freedom of GR, 
there is an additional massive one hidden in the trace sector; contracting with $g^{ab}$ gives
\begin{equation}
f'(R) R-2f(R)+3\Box f'(R)=0.
\label{EqfRTrace}
\end{equation}
In GR, where $f(R)=R-2\Lambda$ and $f''(R)=0$, this scalar equation reduces to a mere \textit{algebraic constraint}. 
In metric $f(R)$-gravity, some would-be boundary structure of GR is no longer dynamically inert, so that Eq~\eqref{EqfRTrace} becomes a \textit{differential equation} for $R$, making $R$ genuinely dynamical: a propagating massive scalar degree of freedom often called the ``scalaron".

The idea behind moving to a Brans-Dicke formalism is to \textit{make this scalar degree of freedom explicit at the level of the action}. Introduce an auxiliary scalar field $\chi$ and write
\begin{equation}
S_{\chi}[g,\chi]
=
\frac{M_P^2}{2}
\int d^4x\sqrt{-g}
\left[
f(\chi)+f'(\chi)(R-\chi)
\right].
\label{EqAuxiliaryAction}
\end{equation}
Varying w.r.t. $\chi$ gives
$
f''(\chi)(R-\chi)=0, 
$
 one obtains $\chi=R$ (on the condition that $f''(\chi)\neq 0$), which recovers Eq.~\eqref{Eqf(R)}. Now define
\begin{equation}
\Phi=f'(\chi),
\qquad
U(\Phi)=\chi(\Phi)\Phi-f(\chi(\Phi)), 
\label{EqLegendre}
\end{equation}
with $U(\Phi)$ the Legendre transform of $f(R)$.
Then, Eq.~\eqref{EqAuxiliaryAction} becomes
\begin{equation}
S^{JF}_{BDT_{0,U}}[g,\Phi]
=
\frac{M_P^2}{2}
\int d^4x\sqrt{-g}
\left[
\Phi R-U(\Phi)
\right],
\label{EqJordanFrame}
\end{equation}
assuming the map 
 $\chi\mapsto f'(\chi)$ is locally invertible. 
  $BDT^{JF}_{0,U}$ abbreviates \textit{Brans-Dicke theory} (cf.~\cite{Duerr2021-DUETIA}) in the Jordan frame, $\Phi$ the Brans-Dicke scalar, 
  $\omega_{BD}=0$ the Brans-Dicke parameter, and $U$ a scalar potential. 
 Here, $\Phi$ is non-minimally coupled to curvature,  has no explicit kinetic term, and its own Euler–Lagrange equation is algebraic, i.e. 
$
U'(\Phi)=R
% -
% \frac{2\omega_{BD}}{\Phi}\Box\Phi
% +
% \frac{\omega_{BD}}{\Phi^2}(\nabla\Phi)^2, 
$. Yet,
   $\Phi$ indirectly influences dynamics through its presence in the modified Einstein equations: 
% varying w.r.t $\Phi$ still gives an algebraic constraint, 
% but the $\Phi$ does show up in the modified Einstein equations. 
\begin{equation}
\Phi G_{ab}
=
% \frac{1}{M_P^2}T_{ab}
% +
% \frac{\omega_{BD}}{\Phi}
% \left(
% \nabla_a\Phi\nabla_b\Phi
% -
% \frac{1}{2}g_{ab}(\nabla\Phi)^2
% \right)
% +
\nabla_a\nabla_b\Phi
-
g_{ab}\Box\Phi
-
\frac{1}{2}g_{ab}U(\Phi).
\label{EqBDMetricEOM}
\end{equation}
%  Equivalently, after taking the trace of Eq.~\eqref{EqBDMetricEOM}
% and combining it with Eq.~\eqref{EqBDScalarEOMRaw}, one obtains
% \begin{equation}
% (2\omega+3)\Box\Phi
% =
% \frac{1}{M_P^2}T
% +
% \Phi U'(\Phi)
% -
% 2U(\Phi),
% \label{EqBDScalarEOM}
% \end{equation}
% where $T=g^{ab}T_{ab}$.
\noindent    Eq.~\eqref{EqJordanFrame} is sufficiently general for this chapter, but the bigger theory landscape is  insightful: generally, BDT's look like 
\begin{equation}
S_{\text{BDT}_{\omega_{BD},U}}[g,\Phi]
=
\frac{M_P^2}{2}
\int d^4x\sqrt{-g}
\left[
\Phi R
-\frac{\omega_{BD}}{\Phi}g^{ab}\nabla_a\Phi\nabla_b\Phi
-U(\Phi)
\right], 
\label{EqBDT}
\end{equation}
which is itself a subclass of \textit{scalar-tensor theories}:
\begin{equation}
S_{\text{ScalarTensor}}[g,\Phi]
=
\int d^4x\sqrt{-g}
\left[
\frac{M_P^2}{2}F(\Phi)R
-\frac{1}{2}Z(\Phi)g^{ab}\nabla_a\Phi\nabla_b\Phi
-V(\Phi)
\right].
\label{EqScalarTensor}
\end{equation}
Scalar-tensor theories admit of (among others) the so-called \textit{Jordan frame} (JF) and  \textit{Einstein frame} (EF) representations. 
Thorne, Lee, and Lightman (\cite*[pp.~3564--5]{thorne1973foundations}) call these ``mathematical representations of a theory", as long as they give the same predictions.   (A ``frame'' is understood in that representational sense; not, for example, as a frame of reference.)
  The use of the EF is to decouple the scalar from curvature, through a conformal transformation with the scalar itself as the conformal factor: 
\begin{align}
\tilde{g}_{ab}=\Phi g_{ab}, 
\qquad
\Phi>0.
\qquad 
% \tilde{\Box} \equiv g^{ab}\nabla_a\nabla_b
\label{EqConformalTransformation}
\end{align}
 Through standard but tedious identities \citep[%Sec.~3.1.1, 
Eqs.~(100)--(104)]{CliftonFerreiraPadillaSkordis2012}, 
\begin{align}
 \sqrt{-g}=\Phi^{-2}\sqrt{-\tilde g},
\qquad
R=\Phi\left[
\tilde R
+3\tilde\Box\ln\Phi
-\frac{3}{2}\tilde g^{ab}\tilde\nabla_a\ln\Phi\,\tilde\nabla_b\ln\Phi
\right], 
\label{EqConformalIdentities}
\end{align}
% so that  
  Eq.~\eqref{EqJordanFrame}  
 (neglecting the total boundary term $\int d^4x\sqrt{-\tilde g}\tilde\Box\ln\Phi$) 
  becomes 
\begin{equation}
S^{EF}_{BDT_{0,U}}[\tilde g,\Phi]
=
\frac{M_P^2}{2}
\int d^4x\sqrt{-\tilde g}
\left[
\tilde R
-\frac{3}{2\Phi^2}
\tilde g^{ab}\tilde\nabla_a\Phi\tilde\nabla_b\Phi
-\frac{U(\Phi)}{\Phi^2}
\right], 
\label{EqEinsteinFramePhi}
\end{equation}
where $\tilde{\nabla}_a$ is built from $\tilde{g}_{ab}$ and 
$\tilde{\Box} \equiv \tilde{g}^{ab}\tilde{\nabla}_a\tilde{\nabla}_b$; for scalars, 
%first derivatives are connection-independent: 
$\tilde{\nabla}_a\Phi=\partial_a\Phi=\nabla_a\Phi$.
%  Varying Eq.~\eqref{EqEinsteinFramePhi} w.r.t. $\tilde{g}_{ab}$ gives
% \begin{equation}
% \tilde G_{ab}
% =
% \frac{3}{2\Phi^2}
% \left(
% \tilde\nabla_a\Phi \tilde\nabla_b\Phi
% -\frac{1}{2}\tilde g_{ab}
% \tilde g^{cd}\tilde\nabla_c\Phi\tilde\nabla_d\Phi
% \right)
% -\frac{1}{2}\tilde g_{ab}\frac{U(\Phi)}{\Phi^2}.
% \label{EqEinsteinFrameMetricEOM}
% \end{equation}
% Varying with respect to $\Phi$ gives
% \begin{equation}
% 3\Phi \tilde\Box \Phi
% -
% 3\tilde g^{ab}\tilde\nabla_a\Phi\tilde\nabla_b\Phi
% -
% \Phi U'(\Phi)
% +
% 2U(\Phi)
% =
% 0.
% \label{EqEinsteinFramePhiEOM}
% \end{equation}
 The scalar and curvature are now decoupled, but $\Phi$'s kinetic term is not canonically normalized. 
 This is solved by field redefinition:
 \begin{equation}
\varphi
=
\sqrt{\frac{3M_P^2}{2}}\ln\Phi,
\qquad
\Phi
=
\exp \left( \sqrt{2/3M_P^2}\,\varphi \right). 
\label{EqCanonicalScalar}
\end{equation}
The EF-action is then
\begin{equation}
S^{EF}_{BDT_{0,U}}[\tilde g,\varphi]
=
\int d^4x\sqrt{-\tilde g}
\left[
\frac{M_P^2}{2}\tilde R
-\frac{1}{2}\tilde g^{ab}\tilde\nabla_a\varphi\tilde\nabla_b\varphi
-V_E(\varphi)
\right], 
\quad 
V_E(\varphi)
=
\frac{M_P^2}{2}
\frac{U(\Phi(\varphi))}{\Phi(\varphi)^2}.
\label{EqEinsteinFrame}
\end{equation}
% for
% \begin{equation}
% V_E(\varphi)
% =
% \frac{M_P^2}{2}
% \frac{U(\Phi(\varphi))}{\Phi(\varphi)^2}.
% \label{EqEinsteinPotential}
% \end{equation} 
Varying Eq.~\eqref{EqEinsteinFrame} w.r.t. $\tilde{g}_{ab}$ gives
\begin{equation}
\tilde G_{ab}
=
\frac{1}{M_P^2}
\left[
\tilde\nabla_a\varphi \tilde\nabla_b\varphi
-\frac{1}{2}\tilde g_{ab}
\tilde g^{cd}\tilde\nabla_c\varphi\tilde\nabla_d\varphi
-\tilde g_{ab}V_E(\varphi)
\right],
\label{EqEinsteinFrameCanonicalMetricEOMAlternative}
\end{equation}
where $\tilde G_{ab}:=\tilde R_{ab}-\frac{1}{2}\tilde g_{ab}\tilde R$. Varying w.r.t. $\varphi$ generically gives a massive Klein-Gordon equation
\begin{equation}
\tilde\Box\varphi
-
\frac{dV_E}{d\varphi}
=
0.
\label{EqEinsteinFrameCanonicalScalarEOM}
\end{equation}
 \noindent 
 Note that any particularities of $f$ are absorbed into the potential. 
For the Starobinsky model, 
% \begin{equation}
% f(R)=R+\alpha R^2,
% \label{EqStarobinskyfR}
% \end{equation}
$
f(R)=R+ R^2/6M^2 
$,
one obtains
\begin{equation}
U(\Phi)=\frac{3M^2}{2}(\Phi-1)^2,
\qquad
V_{\text{plateau}}(\varphi)
=
\frac{3}{4} M_P^2 M^2 
\left[
1-\exp\left(-\sqrt{2/3M_P^2}\,\varphi\right)
\right]^2.
\label{EqStarobinskyPotentialM} 
\end{equation}
 The functional form of this potential is suited
  for inflation:  a flat plateau at large $\varphi$, where inflation can take place, and a stable minimum at $\varphi=0$, around which inflation ends and reheating begins. 
  It also fixes slow-roll observables (\cite[\S4.1]{Inflationaris}).

All transformations above are legitimate only on domains where they are \textit{smooth} and \textit{invertible}: $f''(R)\neq 0$ for the Legendre map, $\Phi=f'(R)>0$ for the conformal transformation, and sufficient regularity to avoid zeros or singularities of the conformal factor. %Boundary terms must be handled consistently. The equivalence claimed here is local and classical; quantum equivalence is a further question. 
 When these conditions are met, the action functionals are \textit{dynamically equivalent}, giving the same equations of motion.  
Where invertibility fails, 
there is no empirical equivalence.

Overall,  
the structural landscape is the scalar-tensor branch of the space of spacetimes: 
\begin{equation}
\text{GR}
\subset
\text{Starobinsky}
\subset
f(R)%\text{-gravity}
\sim
(BDT^{JF}_{0,U}
\leftrightarrow
BDT^{EF}_{0,U})
\subset
BDT_{\omega_{BD},U}
\subset
\text{scalar--tensor theories},
\label{EqLandscape} \nonumber
\end{equation}
where $\subset$ means ``is a special case of'', $\sim$ means ``is dynamically equivalent to'' (not necessarily identical), and $\leftrightarrow$ means ``mathematically equivalent representations".

% \clearpage
\section{The conventionalist temptation: physicists and philosophers traversing the spacetime/matter-dichotomy in $f(R)$-gravity}\label{sec:reviewconvtemptation}

Within the physics literature, the Jordan and Einstein frames are often treated as mathematically equivalent but interpretively treacherous, and there are various expressions of the pull of conventionalism about the spacetime/matter-dichotomy. 
For example,  \citet[pp.~142--143]{Quiros2019ScalarTensorTheories} speaks of the ``conformal frames conundrum'', which is ``perhaps, one of the oldest controversies concerning scalar-tensor theories of gravity". 
The conundrum is whether the two frames are physically equivalent and, if not, which one is the ``physical'' one.  
   $f(R)$-gravity, Eq.~\eqref{Eqf(R)}, seemingly concerns spacetime only, being built out of metric quantities; moreover, 
the JF scalar field, Eq.~\eqref{EqJordanFrame}, is non-dynamical and non-minimally coupled to curvature, so that gravity is not ``exclusively a curvature effect'' but partly due to a ``non-geometric'' scalar (\textit{ibid.}, p.~6);  
finally, that scalar appears in the EF, Eq.~\eqref{EqEinsteinFrame}, as an ordinary-looking dynamical matter field with a canonical kinetic term and a (non-polynomial) potential. 
Thus, scalar-tensor theory is a ``non-fully geometrical, metric theory of gravity" (\textit{ibid.}, p.~6).  
 
Following \citet{Dicke1962MachPrinciple}, many take  conformal transformations as position-dependent changes of unit: the physical laws should be invariant under such changes, while quantities such as curvature need not be. 
Apparent inequivalences come from comparing dimensionful quantities or interpreting EF variables with JF rods and clocks (see \cite{DeruelleSasaki2011ConformalEquivalence,PostmaVolponi2014Equivalence,DomenechSasaki2016ConformalFrames}). These authors often write ``matter" in scare quotes; 
Mukhanov (\citeyear[p.~257]{Mukhanov2005PhysicalFoundations}) even uses scare quotes for the ``scalar" field $\phi$.

   A more conventionalist line identifies a conceptual problem (``physical interpretation"). 
 Where \citet[p.~410]{SotiriouLiberatiFaraoni2008NoProgress} state the equivalence claim in strong terms: ``the way we have just introduced [JF and EF] should make it crystal clear that they are just alternative, but physically equivalent, representations of the same theory.''  
 % Yet, they note that the stress-energy tensor in EF is ``not covariantly conserved, unless it describes conformally invariant matter with vanishing trace,'' and that the usual tracking of position-dependent units is ` `most often forgotten in the literature,'' thereby ``leading to the study of a different theory.'' 
 Furthermore, ``the concept of vacuum versus non-vacuum, or of `matter field' versus `gravitational field' is representation-dependent.'' 
  \citet{FaraoniGunzigNardone1999ConformalTransformations} write that ``[t]he two conformal frames, however, are substantially different'', but they do not say why (!), and proceed only with a review of papers in favour of either the EF- or JF-formulation.

Philosophers have paid comparatively little attention to the conformal frames conundrum,  but three works stand out. 
First, \citet{Weinstein1996} anticipates today's  situation: standard criteria for identifying spacetime structure can come apart once no single metric can coordinate geodesic motion and  matter coupling. %standard criteria for identifying spacetime structure can come apart once geodesic motion, matter coupling, and the form of the Einstein equation 
% are no longer coordinated by a single metric. 
 For Brans-Dicke theory (he does not mention $f(R)$), the JF and EF are  ``observationally indistinguishable" but ``appear to have different ontologies", explicated in terms of geometric quantities and matter, respectively (\textit{ibid.}, p.~S68)---a conceptual conflict (iii). 
He outlines (without resolving), three broad interpretative options, which I take to be summarized as: (a) privilege the metric that fulfills the gravitational-variational role in the pure gravity sector, 
or (b) privilege the metric selected on criteria to do with matter coupling, stress-energy conservation, propagation, and/or geodesic motion, %appearing in the GR formulation (the one entering the Einstein tensor), or (b)  privilege the one governing matter propagation of $\varphi$, 
 or (c) deny there is only one true metric.

Second, 
\citet{Duerr2021-DUETIA} grants formal correspondence between $f(R)$-gravity and $BDT_{0,U}$ as well as between the representations $BDT^{JF}_{0,U}$ and $BDT^{EF}_{0,U}$---indeed, Dürr argues these are theoretically equivalent. 
Yet, Dürr rejects ontological equivalence on grounds of \textit{parsimony}: 
  $\Phi=f'(R)$ is reducible to the metric (and, for Palatini $f(R)$-gravity, reducible to the metric-cum-ordinary-matter), and should therefore  not be taken ontologically seriously (more on this in \S\ref{sec:againstparsimony}).   
  % Thus, there is  no threat to the spacetime/matter-dichotomy. %, since the EF is not ontologically viable, although it may be useful representationally.  

  Third,  \citet{FerreiroFleurenMartens} develop explicit supra-theoretic criteria (partly functional, partly structural) for matter and spacetime for  metric field theories. 
  They test these successfully in the familiar waters of   hydrodynamics and electrodynamics, 
and then apply them to scalar-tensor theories. 
They find that the spacetime/matter-dichotomy (detailing out what I called the GR textbook coordination) fails to give a frame-invariant classification of the scalaron: 
the JF-scalar represents neither matter nor spacetime, whereas the EF-scalar represents matter but not spacetime (and  in the hybrid case of EF-quantities with stress-energy is conserved w.r.t. the JF-derivative, the scalar represents spacetime but not matter). 
They then distinguish five interpretive responses  (\cite{Weinstein1996} can be read as a precursor): 
% to the apparent breakdown of the dichotomy:
\begin{enumerate} 
\item \textit{\underline{Spacetime Functionalism}}: spacetime counts as whatever realizes the spacetime role (open to multiple realization).
 \item \textit{\underline{Structural Common Core}}: commit only to the  structure invariant across frames.
\item \textit{\underline{Quietism}}: refrain from adjudicating    frame-variant questions.
\item \textit{\underline{Discrimination}}: privilege  one frame as ontologically perspicuous and understand the non-privileged frame   either
\begin{enumerate} 
    \item \textit{\underline{Fundamentalism}}:   as real but associated with the privileged frame;
    or
    \item \textit{\underline{Eliminativism}}: as  ontologically pathological---to be excised from the ontology. 
\end{enumerate}
\item \textit{\underline{Strong Breakdown}}: abandon the strict spacetime--matter dichotomy, either 
\begin{enumerate} 
    \item \textit{\underline{Conventionalism}}: by keeping the concepts but denying  their application latches on to a matter  of synthetic fact; 
    or
    \item \textit{\underline{Non-Applicability}}: by banishing spacetime and matter as appropriate conceptual categories altogether. %(at least in this context).
\end{enumerate}
 \end{enumerate}
\citeauthor{FerreiroFleurenMartens}  themselves prefer (5b), replacing the spacetime/matter-dichotomy with a gravity/non-gravity-dichotomy. 
Indeed, conceptualizing ``gravitational  field" as coupling only to matter through the matter sector's total stress-energy, without directly coupling to internal structure, the scalaron comes out gravitational in both JF and EF formulations.

In \S\ref{sec:conventionalistbreaksdown}, I argue that matter coupling heuristics  undermine the conventionalist  requirement (ii); there, I will gladly support (4a), namely Discrimination and Fundamentalism (about the conformal frames conundrum!). %taking into account  non-scalaron matter fields. %(\textcolor{red}{funny or too much?})
  First, however, I analyze Dürr's arguments, which amounts to  Discriminatory-Eliminativism (4b), which is crucially different in that it does not involve non-scalaron matter fields. 
% Yet, 

% \clearpage
 \section{Against parsimony}\label{sec:againstparsimony}

 \citet{Duerr2021-DUETIA} identifies another threat to the spacetime/matter-dichotomy, not at the level of the conformal frames conundrum (i.e., \textit{within} $BDT_{0,U}$), but at the level of equivalence of $f(R)$-gravity to $BDT_{0,U}$ (regardless of which frame). 
That is, by moving from the original action \eqref{Eqf(R)} 
to the action with (redefined) auxiliary scalar~\eqref{EqJordanFrame}.

 In Eq.~\eqref{EqJordanFrame}, $\Phi$ is an independently specifiable field apart from the metric. 
 Nonetheless, it is \textit{dynamically reducible} to it: 
 given a solution for $g_{ab}$, this fixes $R$, which in turn fixes $\Phi$. 
 Yet, a given solution $\Phi$ does not fix $g_{ab}$. 
  (I read the intended point not merely as saying that the scalar is correlated with the metric in the actual world,
   but that the correlation holds across all dynamically admissible models.) 
   Dürr then draws this dynamical reducibility into an epistemic, a semantic, and an ontological realm. 
 
 Epistemically, dynamical reducibility is taken to imply \textit{epistemic reducibility}:  
\begin{quote}
    A solution $g$ of the
$BDT_{0,U}$   field equations fully determines $\phi$. In this sense, we'll say
that the Brans-Dicke scalar is dynamically reducible to the metric.
Its dynamical reducibility curtails the scalar's epistemic relevance:
knowledge of the metric exhausts knowledge of the scalar, but not
vice versa. In this sense, the scalar is epistemically reducible vis-à-vis
the metric.
\citep[p.~14, \textit{original emphasis}]{Duerr2021-DUETIA}  
\end{quote}
\noindent Semantically, Dürr concludes from this     the scalar's  ``negligibility \textit{for theory individuation}'' (\textit{ibid.}, p.~14):  %\citep[p.~14]{Duerr2021-DUETIA}: 
$f(R)$-gravity and $BDT_{0,U}$ should
 not count as distinct theories merely because one contains an additional variable whose value is already fixed by other variables. 

Ontologically, the above is taken to support ``ontological reducibility'' (\textit{ibid.}, p.~22)): 
if the scalar is fixed by the metric, then one need not include it in the fundamental ontology:%---``Parsimony counsels such an eliminative strategy'' \citep[p.~22]{Duerr2021-DUETIA}. 
 \begin{quote}
    For $f(R)$-gravity, one can dynamically reduce its BDT-formulation's
scalar degree of freedom to more fundamental degrees of
freedom (see \S3.1). That is, other degrees of freedom completely
determine it (but not vice versa): one needn't postulate an autonomous
dynamics for the scalar. Its dynamical reducibility suggests
the scalar's \textit{ontological} reducibility (or elimination): rather than
regarding it a self-standing entity, it's plausible to excise it from
one's fundamental ontology; the scalar, then, is nothing but the
(more fundamental) degree(s) of freedom to which it's dynamically
reducible. Parsimony counsels such an eliminative strategy. 
\citep[p.~22, \textit{original emphasis}]{Duerr2021-DUETIA} 
\end{quote}

\noindent 
 I embrace Dürr's semantic point that 
 $\Phi$ does not discriminate between physically admissible models once the metric has already been fully specified, and, as such, $f(R)$-gravity and $BDT_{0,U}$ should not be treated as distinct theories 
 merely because the latter contains an additional variable---variable-counting, after all, is not theory-counting.
%  Of course I agree that the scalar does not discriminate between physically admissible models once the metric has already been fully specified: 
% there is no pair of dynamically admissible models that agree on $g_{ab}$ while differing in $\Phi$. 
% I embrace Dürr's semantic argument against individuating theories by merely counting variables. 
 
%\textcolor{red}{ 
 However, although they are much repeated throughout the paper, the inferences from dynamical reducibility to epistemic and ontological reducibility remain, at best,   enthymenes. 
Indeed, to read off ontology suggested by a particular choice of variables can be misleading. 
\citet{KarananasShaposhnikovZell2022FieldRedefinitions} make this vivid %via a toy model, and again in Higgs inflation, by showing how field redefinitions remove apparently problematic structures without changing the invariant physical content: what looked like an additional interaction or a low cutoff was an artifact of variables, not a new degree of freedom. 
 via a toy model that begins with an apparently interacting action containing derivative mixings, 
 \begin{equation}
S = -\frac{1}{2}\int d^4x \left[
(\partial_a \phi_1)^2
+(\partial_a \phi_2)^2
+\frac{2\phi_2}{\tilde{\Lambda}}\partial_a\phi_1\partial^a\phi_2
+\frac{c}{\tilde{\Lambda}^2}\phi_2^2\partial_a\phi_2\partial^a\phi_2
\right].
\end{equation}
But through
 the field redefinitions
 \begin{equation}
\chi_1=\phi_1+\phi_2^2/(2\tilde{\Lambda}),
\qquad
\chi_2=\int^{\phi_2}d\phi\sqrt{1+(c-1)\phi^2/\tilde{\Lambda}^2}
\end{equation} 
this reduces to the simple free form 
\begin{equation}
    S=-\frac12\int d^4x[(\partial\chi_1)^2+(\partial\chi_2)^2]. 
\end{equation} 

The moral is the manifest presence of interactions, cutoffs, or field-content %(the authors point to misconceptions about the Higgs inflation cut-off scale) 
  % may be artifacts of  poor choices of variables. 
  % The manifest presence of interactions, cutoffs, or field content 
    depends on the choice of variables (the authors clarify misconceptions about the Higgs inflation cut-off scale) . 
  %   , 
  %   and does not establish that the variables in which the dynamics appears simplest are ontologically privileged; dynamically perspicuous variables may be calculational tools, while less simple variables refer more directly to physical subsystems.
  % %} 
     On the one hand, this supports Dürr's  caution to treat the Brans-Dicke scalar as ontologically \textit{additional} merely because it manifestly appears. 
  On the other hand, this cuts both ways, and increases the burden on Dürr's eliminativism: 
  to justify eliminating $\Phi$, 
  Dürr must show not just that it is introduced as some misleading reparametrization, 
  but that it fails to correspond to an invariant degree of freedom. 
  That is, Dürr must establish the stronger claim that no invariant physical content is lost by eliminating $\Phi$.

 I argue much would be lost: eliminating $\Phi$ eliminates $f(R)$-gravity as a viable theory beyond GR. 
 The point is that whereas $\Phi$ is indispensable for the Brans-Dicke formulation, it does not thereby mark an \textit{additional} autonomous physical ingredient, as Dürr claims. 
The Legendre transformation \eqref{EqLegendre} is crucial here. 
A Legendre transformation can introduce new variables that make structure explicit without thereby multiplying physical content. 
 Most famously, the  Legendre map from configuration space to phase space, $TQ\to T^*Q$, from a Lagrangian $(q,\dot q)$ formulation to a Hamiltonian $(q,p)$ formulation,  
 introduces generalized momenta $p$ as independent coordinates on phase space. 
 but this does not mean Hamiltonian mechanics postulates additional autonomous physics  over and above Lagrangian mechanics. 
  A possible wrinkle could be the breaking of hyperregularity conditions \citep{Barrett2019-BAREAI-10}, or, in the singular case, when   constraint structure is not properly controlled to account for mapping many $\dot{q}$'s to the same $p$ \citep{Gomes2024-GOMHTC,Bradley2025Irregular}.

  The same applies to the passage from $f(R)$-gravity to $BDT_{0,U}$: $\Phi$ is introduced as a Legendre transform of $f(R)$. 
The analogous regularity conditions are met precisely by $f''(R)\neq 0$ so that $R\mapsto \Phi=f'(R)$ is locally invertible, and also, for EF, by the domain restrictions $\Phi>0$. 
Thus, the new variable merely reorganizes the same dynamics; 
 one does not ``postulate" an autonomous dynamics for $\Phi$. 
Dürr merely shows that $\Phi$ is ``not extra", 
 but not that it is ontologically eliminable.

 In fact, Dürr's parsimony argument can be made without ever looking at Brans-Dicke theories. 
  The trace sector of the modified Einstein equations \eqref{EqfRTrace} \textit{isolates} only the massive scalar sector; 
  it does not determine the two massless modes of standard GR.  
  Full knowledge of $g_{ab}$ determines everything about this scalar sector; 
  yet, full knowledge of the scalar sector does not fix $g_{ab}$! 
  The trace/traceless decomposition makes this explicit: 
  the trace part of Eq.~\eqref{EqfREOM} is Eq.~\eqref{EqfRTrace}, 
and the traceless part   is their difference
\begin{equation}
f'(R)\left(R_{ab}-\frac{1}{4}Rg_{ab}\right)
-
\left(
\nabla_a\nabla_b f'(R)
-\frac{1}{4}g_{ab}\Box f'(R)
\right)
=0, 
\end{equation}
which governs the remaining information about the metric. Thus knowledge of $R$, or equivalently $f'(R)$ (given invertibility), gives all knowledge about the scalar sector 
but not about the traceless metric sector.  
 Following Dürr's reasoning, 
 one should now conclude  the ``ontological reducibility" or ``elimination" of the trace sector. 
 % the autonomous dynamics belongs to the massive gravitational degree of freedom already present in the $f(R)$-theory, namely in the trace of the modified Einstein equations \eqref{EqfRTrace}. 
  But in this context one sees why that is not correct reasoning: 
  the metric $g_{ab}$ is simply the collective variable containing all propagating degrees of freedom, namely the two massless modes of GR and the additional massive scalar mode.  
  The \textit{raison d'être} of the Legendre transformation is simply to make this additional mode explicit: 
   parsimony of variables is not parsimony of degrees of freedom.

\section{Idealization and approximation} \label{sec:idealapprox}

My use of `approximation' and `idealization' are not intended to capture some consensus usage from their vast respective literatures; 
rather, I use them in precise senses borrowed from  \citep{Norton2012Idealization}: 
\begin{itemize}
    \item An \textit{\underline{approximation}} is an inexact description of the same target system. 
    \item An \textit{\underline{idealization}}, by contrast, introduces a distinct system (real or fictional), 
some of whose properties provide an inexact description of the target. 
\end{itemize}
Norton illustrates the difference with a simple model: the velocity $v(t)$ of a unit mass falling in a weakly resisting medium  satisfies 
\begin{equation}\label{EqFallDiffEq}
\frac{dv(t)}{dt}= g - k v(t),
\end{equation}
so that, for $v(0)=0$, the solution is
\begin{equation}\label{EqApproximatedFall}
v(t)
=
\frac{g}{k}\left[1-\exp\left(-k t\right)\right]
=
g t-\frac{g k t^2}{2}+\frac{g k^2 t^3}{6}-\cdots .
\end{equation}
For small $k$ and early times, one may of course \textit{approximate} the motion by $v(t)\approx gt$. 

Yet, what is also often done is that one introduces a different system: a body falling in a vacuum. This is an \textit{idealized} system, where 
\begin{equation}\label{EqIdealizedFall}
     v(t)=gt 
\end{equation} 
holds \textit{exactly}. 
Now, Norton indicates a danger: it is easy to mistake exact features of the idealized system \eqref{EqIdealizedFall} for exact features of the original target system, namely the falling body itself. 
In particular, the danger here would be to treat the exact law of the surrogate vacuum system, $v(t)=gt$, as though it were an exact law of the real resisted fall, rather than an approximation valid only when the resistance term is negligible.  

% \textcolor{red}{Maybe this whole discussion later}. 
I argue this exact issue arises in the putative case of conventionalism about the spacetime/matter-dichotomy for $f(R)$-theories: 
pure gravity is mistaken for the full target system, whereas  
really it is an idealized model in which all sectors except the gravitational one have been idealized away. 
%:  $
% S_{\mathrm{f(R)}}[g_{ab}]
% =
% S^{\mathrm{JF}}_{\mathrm{grav}}[g_{ab},\Phi]
% =
% S^{\mathrm{EF}}_{\mathrm{grav}}[\tilde g_{ab},\phi] 
% $. 
 %  Most of what I will argue is about $f(R)$-theories and their equivalent $BDT_{0,U}$ formulation. 
% As such, the cosmological applications are not the main focus. 
% But it is important for two reasons. 
% First, because they motivate that this is not an exercise into a mere formal extension of GR, but one that is taken seriously in physics communities, in particular for modeling inflation and dynamical dark energy.  
% Second, more importantly, the inflation case motivates the idealization of the $f(R)$-Lagrangian, because during inflation the inflaton dynamics is hypothesized to be \textit{so} dominant that contributions of any other fields can be safely neglected. 

\section{Conventionalism through reifying idealized modeling practices: completing the gravitational/inflationary sector with the Higgs sector} \label{sec:conventionalistbreaksdown}

I argue the formal equivalence on which the conventionalist argument relies holds only for an idealized sector of the theory, in this case the pure gravity sector. 
The main reason for this is that  equivalence between pure-gravity sectors does not by itself determine how those sectors are to be extended by matter: 
pure-gravity-sector equalities typically fail to extend to their completed sectors. 
Once one restores matter that is known to exist, the equality between the corresponding naturally-coupled models no longer follows: the idealized equivalence collapses into mere approximation. 

Similar observations -- although rare  and often serving other goals --  
are sometimes found in the physics literature, (e.g.~\cite{KalloshLinde2013UniversalityClass,EmaMukaidaNakayama2022ScalarFieldCouplings,Rigouzzo:2025hza,Rigouzzo:2026ycb,KarananasShaposhnikovZell2022FieldRedefinitions}). %EmaMukaidaNakayama2022ScalarFieldCouplings, 
Here, the goal is to clarify that what is going on concerns the distinction between approximations and idealizations; additionally, I point at empirical consequences of the heuristics of matter coupling, particularly canonical normalization and minimal coupling, and their interplay with idealization. 

Distinguish two \textit{moments at which} the added-sector form may change:
\begin{itemize}
    \item[(A)] during the \underline{Legendre transformation} \eqref{EqLegendre} from the $f(R)$ action to $BDT_{0,U}^{JF}$;
    \item[(B)] during the \underline{conformal-transformation-cum-field-redefinition} 
    \eqref{EqConformalTransformation}--\eqref{EqEinsteinFrame}
    from $BDT_{0,U}^{JF}$ to $BDT_{0,U}^{EF}$.
\end{itemize}
\noindent Then, there are two  
\textit{ways in which} the extension can fail to preserve equivalence, namely when the formulations differ in terms of: 
\begin{itemize}
    \item[(I)] \underline{Canonical normalization}: for a scalar field $\chi$, the kinetic term has the form
    $-\frac{1}{2}g^{ab}\partial_a\chi\partial_b\chi$, without a field-dependent prefactor or derivative mixing.
    \item[(II)] \underline{Scalaron-minimal coupling}: an independently specified matter action $S_m[\bar{g},\psi]$ depends on the original sector only through the metric $\bar{g}_{ab}$ designated in that formulation, without explicit curvature or scalaron dependence.
\end{itemize}
 \noindent In principle, these are common but defeasible theory-building heuristics---they are not necessary  to count as `matter', nor to be physically admissible.  
 matter equations depend on geometry \textit{only} by coupling directly  to the metric,  not (or only indirectly) to curvature terms such as $\phi^2R$.
 Moreover, one can depart from (II) in two ways: by direct (non-minimal) curvature coupling or by scalaron dependence. 
  By ``added matter'' I mean here a sector independently identified as matter, such as the Standard Model Higgs, rather than the scalaron $\Phi$. 
  In the broader scalar–tensor literature, explicit scalaron-dependence is often called \textit{non-minimal coupling} (to gravity), but this terminology presupposes that the scalaron belongs to the spacetime sector---precisely what we  cannot assume here.  
%This is a bookkeeping distinction between the original and added sectors, not an answer to whether the scalaron is ultimately matter-like or spacetime-like.

First, consider (A). 
 Whether an added field is canonically or non-canonically normalized -- that is, (A-I) -- the Legendre transformation \eqref{EqLegendre} is safe: the map only cares about the $R$-dependence in the gravitational sector, not about kinetic terms of other fields. 
  Likewise, if the added  action has the form $S_m[g,\psi]$ without explicit curvature dependence -- that is, (A-II) -- the Legendre transformation 
   from $f(R)$ gravity 
  to $BDT^{JF}_{0,U}$ 
  leaves that sector untouched. 
Were the original action to contain, instead,  a direct coupling $\xi h^2R$, for example, 
the coefficient of $R$ would become $\Phi+\xi h^2$, 
no longer just $\Phi$. 
 Therefore, the ``pure gravity" idealization of the  $f(R)\sim BDT^{JF}_{0,U}$-equivalence  is not a theorem about arbitrary matter-curvature couplings.

Second, moment (B), the conformal transformation.  
It may be best to make the point in a concrete case: add matter that we know exists, namely the Higgs field (the argument works for other non-conformally invariant fields, too, but  the Higgs is relatively simple and experimentally well confirmed; also, note that this is \textit{not} Higgs inflation, but $f(R)$-inflation with the standard Higgs field).  %is not intended to drive inflation, and its contribution to the inflationary background dynamics may be safely ignored for many purposes. ).  
 This is a complex $SU(2)$ doublet, 
 usually minimally coupled as
$\mathcal{L}_H[g,H]
=
(D_aH)^\dagger D^a H
+
\mu^2 H^\dagger H
-
\lambda (H^\dagger H)^2 .
$
 For the present purpose it is harmless to work in unitary gauge (suppress gauge and Yukawa interactions via $H^{\dagger}(x)=(0,h(x)/\sqrt{2}$)) since we are only interested in the transformation properties of its projection on spacetime, viz. a real scalar $h(x)$ with
\begin{equation}
S_H[g,h]
=
\int d^4x\sqrt{-g}
\left[
-\frac12 g^{ab}\nabla_a h\nabla_b h
-
V_H(h)
\right],
\qquad
V_H(h)
=
\frac{\lambda}{4}h^4
-
\frac12\mu^2h^2  .
\end{equation}
  \noindent Thus, we begin with a canonically normalized Higgs action in the standard ``added-sector" form  called  minimal-scalaron coupling above. 
   The JF-completion is
\begin{equation}
S^{JF}_{\mathrm{grav}+H}[g,\Phi,h]
=
S^{JF}_{\mathrm{grav}}[g,\Phi]
+
\int d^4x\sqrt{-g}
\left[
-\frac{1}{2}g^{ab}\nabla_a h\nabla_b h
-
V_H(h)
\right].
\end{equation}
After the conformal transformation \eqref{EqConformalTransformation}, its Higgs sector becomes
\begin{equation}\label{EqHiggsSectorAfterConformal}
S^{EF,\mathrm{image}}_{H}[\tilde g,\Phi,h]
=
\int d^4x\sqrt{-\tilde g}
\left[
-\frac{1}{2}\Phi^{-1}\tilde g^{ab}
\tilde{\nabla}_a h\tilde{\nabla}_b h
-
\Phi^{-2}V_H(h)
\right].
\end{equation}
After canonical normalization of $\Phi$, as in Eq.~\eqref{EqCanonicalScalar}: 
\begin{equation}\label{EqHiggsEinsteinFrameImage}
S^{EF,\mathrm{image}}_{H}[\tilde g,\varphi,h]
=
\int d^4x\sqrt{-\tilde g}
\left[
-\frac{1}{2}\exp\left(-\beta\varphi\right)
\tilde g^{ab}\tilde{\nabla}_a h\tilde{\nabla}_b h
-
\exp\left(-2\beta\varphi \right) V_H(h)
\right].
\end{equation}
This  is \textit{not} the theory obtained by adding a \textit{canonical} and \textit{scalaron-minimally-coupled} Higgs directly to the EF. 
  \textit{That} theory would be 
\begin{equation}  \label{EqHiggsEinsteinFrameMin}
S^{EF,\mathrm{std}}_{H}[\tilde g,h]
=
\int d^4x\sqrt{-\tilde g}
\left[
-\frac{1}{2}\tilde g^{ab}\tilde{\nabla}_a h\tilde{\nabla}_b h
-
V_H(h)
\right].
\end{equation} 
\noindent Thus, a canonically-normalized and scalaron-minimally-coupled Higgs  becomes non-canonical in the EF: case (B-I). Its kinetic term and potential also become (non-scalaron-minimally) scalaron-dependent: case (B-II).  
The transformed Higgs remains \textit{minimally coupled} in the curvature sense, in that it retains its standard form relative to the composite  metric
$g^{m}_{ab}=\Phi^{-1}\tilde g_{ab}=g_{ab}$. 
 But it did become \textit{non-scalaron-minimally coupled} to $\Phi$. 
% (Likewise, and conversely, a Higgs field added canonically and minimal-scalaron in the EF is not the image of a canonical Higgs in the JF.)
 Hence, 
\begin{equation}
S^{EF,\mathrm{image}}_{\mathrm{grav}+H}[\tilde g,\varphi,h]
\not\equiv
S^{EF,\mathrm{std}}_{\mathrm{grav}+H}[\tilde g,\varphi,h].
\end{equation} 

Likewise, and conversely, a canonical, scalaron-independent Higgs action appended directly to the EF would be
\begin{equation}\label{EqHiggsEinsteinFrameMin}
S^{EF,\mathrm{std}}_{H}[\tilde g,h]
=
\int d^4x\sqrt{-\tilde g}
\left[
-\frac{1}{2}\tilde g^{ab}\tilde{\nabla}_a h\tilde{\nabla}_b h
-
V_H(h)
\right].
\end{equation}
This is not the conformal image of the JF extension:
\begin{equation}
S^{EF,\mathrm{image}}_{\mathrm{grav}+H}[\tilde g,\varphi,h]
\neq
S^{EF,\mathrm{std}}_{\mathrm{grav}+H}[\tilde g,\varphi,h].
\end{equation}
Therefore, exact equality between the idealized ``pure gravity"  JF and EF  sectors  does not imply exact equality between their minimal-scalaron-coupled matter-completions:  
adding a canonical, standard minimal-scalaron coupled Higgs sector in a chosen frame does not commute with the  frame transformation.  

Nor can the difference be removed by simply rescaling  $h$. 
 Not only is any local field transformation always permissible, allowing one to rescale a matter field however one wants, it also does not help. 
One might try  the usual conformal rescaling of a scalar field, namely 
\begin{equation}
h=\Phi^{1/2}\tilde h .
\end{equation}
Using $\Phi=\exp\left(\beta\varphi\right)$, this gives
\begin{equation}
\Phi^{-1/2}\tilde{\nabla}_a h
=
\tilde{\nabla}_a\tilde h
+
\frac{\beta}{2}\tilde h\,\tilde{\nabla}_a\varphi .
\end{equation}
Substituting this into the kinetic term of Eq.~\eqref{EqHiggsSectorAfterLegendre} yields
\begin{equation}
-\frac{1}{2}\Phi^{-1}
\tilde g^{ab}\tilde{\nabla}_a h\tilde{\nabla}_b h
=
-\frac{1}{2}\tilde g^{ab}
\left(
\tilde{\nabla}_a\tilde h
+
\frac{\beta}{2}\tilde h\,\tilde{\nabla}_a\varphi
\right)
\left(
\tilde{\nabla}_b\tilde h
+
\frac{\beta}{2}\tilde h\,\tilde{\nabla}_b\varphi
\right).
\end{equation}
The rescaling therefore removes the explicit prefactor $\Phi^{-1}$ but only by introducing derivative couplings between $h$ and $\Phi$.  
 Also the potential does not recover its original form:
\begin{equation}
\Phi^{-2}V_H(\Phi^{1/2}\tilde h)
=
\frac{\lambda}{4}\tilde h^4
-
\frac{1}{2}\mu^2\Phi^{-1}\tilde h^2.
\end{equation}
Whereas the quartic term is invariant under the rescaling,   the Higgs mass term remains explicitly coupled to $\Phi$. 
 Indeed, one can infer that no rescaling simultaneously restores the canonical kinetic term and the original Higgs potential.%} % The scalaron reappears either in the kinetic sector, in derivative mixings, or in the potential. 
 % Besides,  one is free to decide to rescale a matter field or not: initially, a local field transformation is always permissible. 

Why might one nevertheless append Eq.~\eqref{EqHiggsEinsteinFrameMin} when starting from the pure EF? 
The motivation involves soft theory-building heuristic: 
if one restricts a scalar sector to fields of mass dimension at most 4, the kinetic prefactor will be constant and can be removed by a field redefinition. 
 A field-dependent factor such as $\exp\left(-\beta\varphi\right)$ instead expands into a tower of higher-dimensional interactions and introduces a  cutoff scale, 
 which 
 lends some support to the standard extension. 
 But  the argument can hardly  be said to be decisive: 
  GR is already non-renormalizable, 
 and, for inflation, inflaton potentials usually contain operators of higher than 4 dimensions. 
In any case, if you allow for a non-canonical kinetic term for the Higgs, one would be hard-pressed to justify excluding a non-canonical kinetic term for the inflaton.  

Direct Higgs-curvature coupling has a different status. Generically, terms like $\xi h^2R$ are allowed (indeed they should be expected to be generated at higher levels even if absent at tree level). 
Hence what is important is the magnitude of the (renormalized) coupling, rather than simply its presence.  
  Again, this is a another soft model-building heuristic, but it is not what distinguishes  Eqs.~\eqref{EqHiggsEinsteinFrameImage} and \eqref{EqHiggsEinsteinFrameMin}.

The anti-conventionalist argument is not that field redefinitions break invariance: if the full JF-formulation, including the matter sector, is transformed consistently, with careful bookkeeping one obtains the \textit{exact} EF-image of the JF (i.e., Eq.~\eqref{EqHiggsEinsteinFrameImage}).   
Nor is it that one should go to jail for coupling matter non-scalaron-minimally---minimal coupling is only the \textit{simplest} way to couple matter (\cite{Butterfield2007-BUTRRC}). 
However, one would only write down the EF-\textit{image}, rather than a straightforward completion,  if one ``already knew" the JF-formulation.  
That is, without the above derivations, one would naturally ask why we use this non-canonical kinetic term for the scalaron and not for other fields. Viewed as the exact image of a specified JF extension, the field-dependent Higgs kinetic term is not mathematically mysterious: its origin is fixed by the conformal transformation. 
But that genealogy does not provide a frame-independent reason for selecting that particular extension when starting from the pure EF sector---that selection requires independent physical or methodological input.    %(The fact that non-canonical structure is manifest in different places in the two frames does not explain why the full theory should contain one canonically normalized scalar and one non-canonically normalized scalar, rather than two canonical scalars.) 

 % That is,  without the above derivations, one would naturally ask why we use this non-canonical kinetic term for the scalaron and not for other fields---a mystery  further enhanced from an effective field theory perspective.  

 Rather, the non-conventional element lies in the fact that 
the manner in which matter is coupled is not (just) a matter of preference:    
  different choices of canonical normalization and scalaron-minimal or non-scalaron-minimal coupling generally define \textit{different} completed theories, and in principle change empirical predictions.  
  Current evidence does not determine the Higgs-scalaron interaction or any direct Higgs-curvature coupling, but, supposedly, there are determinate (yet currently unknown) scale-dependent values for these couplings.   
     For inflation, sufficiently large couplings affect fluctuation amplitudes, the scalar spectral index, and the tensor-to-scalar ratio, 
  and indirectly affect isocurvature production, the reheating mechanism, and back-reaction dynamics \citep{Rigouzzo:2025hza,GrahamRacco2025IsocurvatureBounds}.  
     For inflation those changes are numerically utterly negligible, which is precisely why the idealized Lagrangian can be used---but negligibility of a difference is not identity.

Perhaps the conventionalist could   insist there are salient differences in the ``would-be matter" and ``would-be spacetime" content  between the exact images of the JF-quantities in the EF and \textit{vice versa}, without privileging either.  
 Does this constitute a genuine underdetermination from which we can informatively conclude that there are no synthetic facts about spacetime and matter in this theory? 
 The answer is either negative, as argued above, or 
 trivial. 
 The threat to non-triviality requirement (iv) arises when the entire interpretation is transported from one frame into the other: every physical magnitude, matter coupling, rod-and-clock behaviour, and representational role on one side is defined as the transformed image of its counterpart on the other side. 
Then, indeed, the two formulations differ only in variables.   But mere reparametrization is always available in field theory: perform some field transformation and insist that the different representations are equally good though conceptually conflicting descriptions.  There is no novel or informative conventionalism here. % about spacetime and matter.
% However, such a conventionalism can be had in any field theory, 
% Thus, at the very least, $f(R)$-gravity or Jordan and Einstein frame formulations do not provide any novel conventionalist argument; indeed, one could insist that two conformally equivalent spacetimes within GR are meaningfully distinct. 
 
Therefore, the conventionalist's dilemma is:
\begin{itemize}
    \item Either: coordinate each frame independently of the other, treating its metric as representing spacetime and completing it with canonically-normalized matter in the standard added-sector scalaron-minimal form,  
    so that
      alternatives are substantively distinct, but no longer formally equivalent (ii).  
    \item Or: transport the whole field and interpretive dictionary, 
    preserving exact equivalence by reparametrising alternatives, but thereby threatening non-triviality   (iv).
\end{itemize}
 \noindent     Insofar as there is a problem   it is an empirical   (not a philosophical) problem, in that both $g_{ab}$ and $\tilde{g}_{ab}$ have a claim on having natural but different features of spatiotemporal representation. 
   Namely, in the JF, $g_{ab}$  is natural because matter is usually taken to be minimally coupled to it, so that it governs rods, clocks, and free fall in a familiar way; 
   in the EF, the gravitational sector takes the Einstein-Hilbert form, and the scalar degree of freedom takes a familiar matter form, making it dynamically more perspicuous. 
   In GR, both these features  can be borne by one and the same metric tensor.  
    But having two natural candidates for spatiotemporality does not imply the choice between them is conventional: at least in principle, there is an empirical resolution.

  \section{Encore}
One might object the whole discussion already grants too much to the old spacetime/matter-dichotomy, inherited from GR textbook coordination. 
For example, \citet{SalimkhaniForthcomingTraditionalConcepts} argues that the apparent collapse of the dichotomy may be engineered by the method used to explicate them: if the relevant criteria are assembled inductively from different practices, one should expect a third category and/or overlap.  
 Why not instead engineer a theory-independent concept of ``spacetime", rather than inductively extracting it from a specifically selected range of theories, and define ``matter"  as its complement, thereby guaranteeing a dichotomy? 
% Why not conceptually engineer ``spacetime" without relying on some specific set of theories, and its negation ``non-spacetime" as synonymous with ``matter", thereby saving the dichotomy? 
Indeed,
\citet{Rynasiewicz1996-RYNAVR} warns that already the pre-relativistic dichotomy between space and matter, or container and contained, is formulated in a conceptual framework foreign to relativity theory. 
Surely, there may be no unique or natural projection of old categories onto the new formalism.

We should heed such skepticism, also for $f(R)$-gravity and Einstein and Jordan frames. 
That being said, it does not follow (nor do Rynasiewicz or Salimkhani claim) %see also Rynasiewicz contribution in this volume? 
that conceptual projection is pointless.  
As \citet[Ch.~1]{Hesse1961} emphasizes already in her early work, empirical concepts are \textit{open-textured} rather than analytically nailed down by necessary and sufficient conditions;  they have an open fringe of meaning that allows them to be revised and transplanted into new empirical domains. 
Indeed, GR textbook coordination of ``spacetime"  and ``matter" seems quite successfully employed, 
even if not in the same way as in pre-relativistic physics, 
and even if sometimes too vaguely or implicitly to satisfy the philosopher. 

Still, such transplantation better be disciplined:  
a full interpretation of $f(R)$-gravity and its $BDT_{0,U}$ formulations should ultimately, I believe, take seriously \textit{Brown's moral} (\cite[Ch.~8]{Brown2005-RBRPRS-2}; \cite[\S4]{Butterfield2007-BUTRRC}). 
Brown's moral says that  
a mathematical structure such as a metric tensor receives chronogeometric significance not by virtue of being a symmetric rank-2 tensor field but, 
 rather, through detailed physical arguments showing \textit{how} rods, clocks, light rays, and matter display the relevant spatiotemporal structure.  
In $f(R)$-gravity this would require a careful study of matter coupling, limiting relations to GR, the  behavior of physical probes, rigorous treatment of the geodesic principle and energy conditions within this formalism, certainly guided by familiarity with GR (cf.~\cite{MarchBPhil}) but logically distinct from it. 
  This chapter followed that  spirit  without settling that larger project of chronogeometric significance but paves the path towards it: involving matter coupling heuristics shows how the multiple spatiotemporal roles of GR's metric tensors come apart in Brans-Dicke theory. 

%\textcolor{red}{ 
 In particular, I showed the narrower point that involving matter coupling heuristics undermines conventionalist arguments against the spacetime/matter-dichotomy that come off the ground because of the idealization of confining one's model to the ``pure gravity" sector only. 
   I do not think such idealizations conflict with conventionalism \textit{per se}, nor that conventionalism \textit{must} always begin with a fully de-idealized, completed theory. 
Rather, I think one should not forget the scope of the argument: equivalence within an idealized pure-gravity sector does not support the corresponding conclusion about a completed, matter-containing theory, because pure-sector equivalence does not uniquely determine equivalent matter-containing extensions. 
If one extends each frame independently, the resulting theories are non-trivially different, but no longer empirically equivalent. 
 If one instead transports the complete matter and interpretive dictionary through the transformation, equivalence is preserved  but at the cost of trivial semantic reconstrual.% 
 
 Perhaps the dichotomy cannot be upheld for other reasons given in this volume, 
but -- absent the unified field theory -- we should not base a claim that spacetime and matter are non-distinct on   idealizations made for pragmatic reasons of modeling practices.

% \clearpage
\setstretch{1.0}
\section*{Acknowledgements}
% \textcolor{red}{Send Frank/Eleanor?/India your pages about}: 
% \begin{itemize}
%     \item  the original Starobinsky stuff  
%     \item  the Legendre transformation argument   
%     \item  the connection with Lagrangian to Hamiltonian 
%     \item  are JF and EF theoretically equivalent?
% \end{itemize}
  % {\footnotesize 
  I am very grateful for conversations with  
 India Bhalla-Ladd, % (IF you ask here to read - maybe, geometric look at this could be an option)
% Jeremy Butterfield (IF you ask him to read - maybe, maybe not: spatiotemporal signifiance of f(R) theories with him could be an option)
Alex Fleuren, 
Antonio Ferreiro, 
% Guy Hetzroni (completeness - only if you mention this) 
Frank Hu, 
Ted Jacobson, %(about f(R) theories) 
 Niels Linnemann, 
Eleanor March, 
John Norton, %(for emails about approx and idealizations, as well as Einstein's Geometry and Experience
Robert Rynasiewicz, 
Kian Salimkhani, 
David Wallace, % (about f(R) theories and about this argument with the Higgs)  
Jim Weatherall, 
  Will Wolf, %(for looking with you at the argument with the Higgs). 
and especially Sebastian Zell. 
I also thank audiences in Washington D.C., Pittsburgh, and Irvine CA, as well as the American Institute of Physics for financial support. % generously provided through the ``Robert H.G.~Helleman Memorial Postdoctoral Fellowship". %, 
%and grant number  ... 
% }

\section*{Statement about GenAI use} 
Generative AI tools were used exclusively for language editing,
including grammar checking, improving clarity and flow, as well as lowering word count. 
AI-generated suggestions were critically reviewed and edited by the author. 
The author takes full responsibility for the final content, including  accuracy, arguments, and conclusions. 

  % \clearpage
%Appendix starts here%
% \part*{Appendices} \addcontentsline{toc}{section}{\protect\numberline{}Appendices}
 % \numberwithin{equation}{part}
 %    \appendix
 %        \numberwithin{equation}{section}

% \section{Appendix: ... } \label{sec:appendix}
% \setstretch{1.0}

\renewcommand{\baselinestretch}{1.0}  
 % \clearpage
 % {\footnotesize 
% }
% \bibliographystyle{apalike}
% \addcontentsline{toc}{section}{Bibliography} 
%  \addcontentsline{toc}{section}{\protect\numberline{}Bibliography}
% \renewcommand\refname{Bibliography}
\begin{refcontext}[sorting=nyt]
% \DeclareNameAlias{sortname}{first-last}
% Define a new name format for the first author
\printbibliography%[]
\end{refcontext}

\end{document}